\documentclass[aps,prb,floatsfix,twocolumn,superscriptaddress,showpacs]{revtex4-1}

\usepackage{dcolumn} 
\usepackage{bm} 
\usepackage{amsthm}
\usepackage{latexsym}
\usepackage{float}
\usepackage{amsfonts}
\usepackage{amsmath}
\usepackage{amssymb}
\usepackage{color,graphicx}

\begin{document}
\title{Fifth-order nonlinear optical response of excitonic states in an InAs quantum dot ensemble measured with 2D spectroscopy}

\author{G.~Moody}
\email{moodyg@jilau1.colorado.edu}
\affiliation{JILA, University of Colorado \& National Institute of Standards and Technology, Boulder CO 80309-0440}
\affiliation{Department of Physics, University of Colorado, Boulder CO 80309-0390}
\author{R.~Singh}
\affiliation{JILA, University of Colorado \& National Institute of Standards and Technology, Boulder CO 80309-0440}
\affiliation{Department of Physics, University of Colorado, Boulder CO 80309-0390}
\author{H.~Li}
\affiliation{JILA, University of Colorado \& National Institute of Standards and Technology, Boulder CO 80309-0440}
\author{I.~A.~Akimov}
\affiliation{Experimentelle Physik 2, Technische Universit$\ddot{\textit{a}}$t Dortmund, D-44221 Dortmund, Germany}
\affiliation{A. F. Ioffe Physical-Technical Institute, Russian Academy of Sciences, 194021 St. Petersburg, Russia}
\author{M.~Bayer}
\affiliation{Experimentelle Physik 2, Technische Universit$\ddot{\textit{a}}$t Dortmund, D-44221 Dortmund, Germany}
\author{D.~Reuter}
\affiliation{Lehrstuhl fuer Angewandte Festkoerperphysik, Ruhr-Universitaet Bochum, Universitaetsstrasse 150, D-44780 Bochum, Germany}
\author{A.~D.~Wieck}
\affiliation{Lehrstuhl fuer Angewandte Festkoerperphysik, Ruhr-Universitaet Bochum, Universitaetsstrasse 150, D-44780 Bochum, Germany}
\author{S.~T.~Cundiff}
\affiliation{JILA, University of Colorado \& National Institute of Standards and Technology, Boulder CO 80309-0440}
\affiliation{Department of Physics, University of Colorado, Boulder CO 80309-0390}

\begin{abstract}
Exciton, trion and biexciton dephasing rates are measured within the inhomogeneous distribution of an InAs quantum dot (QD) ensemble using two-dimensional Fourier-transform spectroscopy.  The dephasing rate of each excitonic state is similar for all QDs in the ensemble and the rates are independent of excitation density.  An additional spectral feature -- too weak to be observed in the time-integrated four-wave mixing signal -- appears at high excitation density and is attributed to the $\chi^{\textrm{(5)}}$ biexcitonic nonlinear response.
\end{abstract}

\date{\today}
\pacs{78.67.Hc, 78.47.jh, 73.21.La}
\maketitle

The optical spectrum of semiconductor quantum dots (QDs) at low temperature is dominated by excitonic features including neutral and charged excitons (trions) and bound or anti-bound two-excitons (biexcitons) \cite{Bayer2003}.  Exciton dephasing times up to nanoseconds\cite{Borri2001,Berry2006} have made semiconductor QDs attractive for applications in quantum information and coherent control, motivating studies investigating dephasing and relaxation mechanisms that can limit the performance of QD-based devices.  The most widely-used technique for studying these properties has been spectrally- and time-resolve photoluminescence (PL) spectroscopies \cite{Brunner1994,Gammon1995,Rodt2002}, in which single QDs must be isolated to overcome significant inhomogeneous line broadening resulting from QD size dispersion.  These experiments have demonstrated the effects of thermal broadening \cite{Gammon1996a,Besombes2001,Ortner2004,Peter2004,Rudin2006} on the exciton homogeneous lineshape and width (inversely proportional to the dephasing time) and on the multi-particle emission energies \cite{Bayer2002a,Schliwa2009}.  Additional insight into dephasing mechanisms and coherent interactions between excitons can be gained by using nonlinear techniques, in which many-body interactions give rise to distinct features in the nonlinear signals\cite{Chemla2001}.  Transient four-wave mixing (FWM) has been particularly suitable for revealing the effects of phonon and inter-exciton scattering on the dephasing rate of excitons, biexcitons and trions even in the presence of strong inhomogeneity \cite{Fan1998,Langbein2004,Borri2007,Cesari2010}.

More recently, optical two-dimensional Fourier-transform spectroscopy \cite{Cundiff2012} (2DFTS) -- an extension of three-pulse FWM -- has been demonstrated as an extremely sensitive tool for investigating coherent excitonic interactions \cite{KasprzakJ.2011,Kasprzak2012,Moody2011b} and incoherent relaxation dynamics \cite{Moody2011} in interfacial GaAs QDs.  2DFTS is advantageous for investigating QD ensembles because of its ability to unfold the coherent response onto two frequency dimensions, separating the homogeneous and inhomogeneous linewidths and clearly isolating different spectral features that would otherwise overlap using one-dimensional techniques.  In this Brief Report, we use 2DFTS to investigate the nonlinear optical response of excitons, trions and biexcitons in an InAs QD ensemble by exploiting the dipole transition selection rules for this system.  2DFTS has not been applied previously to the study of InAs QDs, which exhibit stronger confinement compared to GaAs QDs, and therefore smaller dipole moments, making them more difficult to study.  By using co- and cross-linearly polarized excitation and detection schemes, we isolate the different excitonic states and measure the emission energy dependence of the linewidths.  By increasing the excitation intensity to drive the system beyond the $\chi^{\textrm{(3)}}$ regime, we find that an additional spectral feature, which is too weak to be observed in the time-integrated FWM signal, appears in the 2D spectrum and is attributed to the $\chi^{\textrm{(5)}}$ biexciton nonlinear response that is radiated in the FWM phase-matched direction.  Furthermore, we find that the linewidths of all excitonic states are independent of excitation density, indicating that excitation-induced dephasing (EID) effects are not important at these excitation levels.

\begin{figure}[h]
\centering
\includegraphics[width=0.95\columnwidth]{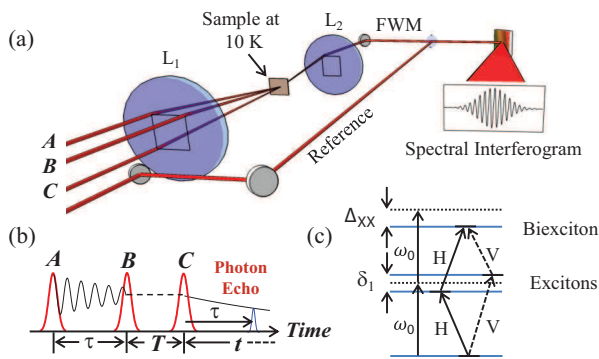}
\caption{(Color online) A schematic of the 2DFTS experimental setup is shown in (a).  Three pulses are focused onto the sample by lens L$_{\textrm{1}}$ to generate a nonlinear signal that is radiated in the phase-matched direction and collimated by lens L$_{\textrm{2}}$.  The signal is heterodyned with a phase-stabilized reference pulse and their interference is spectrally resolved.  Interferograms are recorded while the delay between the first two pulses, $\tau$, is incrementally stepped.  The data is Fourier transformed with respect to $\tau$ to unfold the complex nonlinear response onto two frequency dimensions.  The pulse time ordering is shown in (b) and an energy level diagram for a neutral QD is shown in (c) for the lowest energy exciton states that are energetically separated by the fine-structure splitting, $\delta_{\textrm{1}}$, and the bound biexciton red-shifted from the two-exciton by the biexciton binding energy, $\Delta_{\textrm{XX}}$.}
\label{fig1}
\end{figure}

The sample investigated consists of 10 quantum-mechanically isolated self-assembled InAs/GaAs QD layers epitaxially grown on a GaAs (001) substrate.  Investigation of an ensemble eliminates any modification to the dielectric environment that occurs from patterning the sample -- as is often done in single QD studies -- which might influence the optical properties.  The sample is thermally annealed post-growth at $900^{\circ}$C for 30 seconds, which affects the emission properties of the ensemble in several ways \cite{Leon1996,Greilich2006}: the ground state distribution is blue-shifted to 1345 meV; the inhomogeneous width is narrowed to 15 meV full-width half-maximum (FWHM); and the ground state-to-wetting layer confinement is decreased to 100 meV.  Impurities unintentionally introduced during growth result in half of the QDs being doped with a hole, determined through single quantum dot studies of similar samples and a quantitative analysis discussed below.  In charge-free QDs, electron-hole exchange couples the exciton spin states, forming two linear orthogonally-polarized states separated by the fine-structure splitting ($\delta_{\textrm{1}}$) for an asymmetric confinement potential \cite{Bayer2002a}.  These states are coupled through confinement-enhanced Coulomb interactions, forming the four-level diamond scheme shown in Fig. \ref{fig1}(c).  Biexciton states are optically excited by two co-linearly polarized pulses, which form a four-particle correlated state that is red-shifted (blue-shifted) from twice the average exciton energy by a positive (negative) biexciton binding energy, $\Delta_{\textrm{XX}}$.  Through auxiliary experiments \cite{Moody2012b} measuring $\delta_{\textrm{1}}$, we find that the linearly-polarized exciton states are aligned along the same crystal axes for all QDs in the ensemble, and we define these axes as H and V.  The sample temperature is held at 10 K.

The 2DFTS technique, shown in the schematic diagram in Fig. \ref{fig1}(a), is based on three-pulse FWM with the addition of interferometric stabilization of the pulse delays (for experimental details see Bristow \textit{et al.}\cite{Bristow2009}).  Briefly, 150-fs pulses tuned to the center of the inhomogeneous distribution with a FWHM of 10 meV are incident on the sample with wave vectors $\textbf{\textit{k}}_{\textit{a}}$, $\textbf{\textit{k}}_{\textit{b}}$ and $\textbf{\textit{k}}_{\textit{c}}$.  The pulses generate a nonlinear signal that is radiated in the phase-matched direction $\textbf{\textit{k}}_{\textit{s}}=-\textbf{\textit{k}}_{\textit{a}}+\textbf{\textit{k}}_{\textit{b}}+\textbf{\textit{k}}_{\textit{c}}$. The signal is heterodyned with a phase-stabilized reference and their interference is spectrally resolved with a 17 $\mu$eV resolution.  The delay between the first two pulses incident on the sample, $\tau$, is stepped with interferometric precision while recording spectral interferograms, and the extracted nonlinear signal is Fourier-transformed with respect to $\tau$ to generate a complex 2D rephasing spectrum.  At each delay, the phases of pulses $A$ and $B$ are toggled by $\pi$ using liquid crystal modulators and phased interferograms are combined appropriately to cancel scatter of the excitation beams along $\textbf{\textit{k}}_{\textit{s}}$.  The excitation intensity can be varied from 5 to 200 W cm$^{-2}$ (0.025 to 1 $\times$ $10^{13}$ photons pulse$^{-1}$ cm$^{-2}$).

\begin{figure}[h]
\centering
\includegraphics[width=0.7\columnwidth]{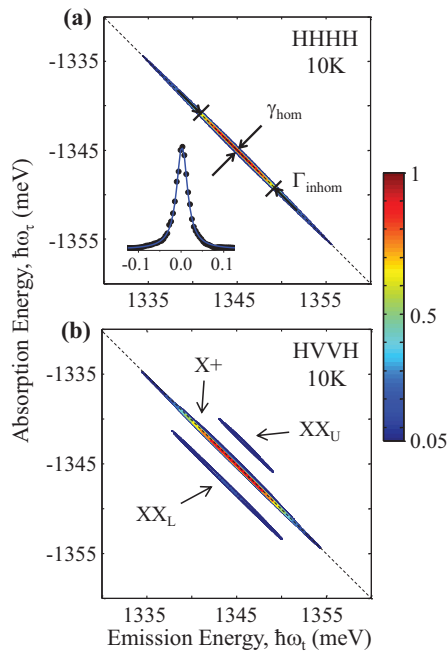}
\caption{(Color online) Normalized rephasing amplitude spectra for co- (HHHH) and cross- (HVVH) linear polarization for an excitation intensity of 200 W cm$^{-2}$ are shown in (a) and (b), respectively.  A cross-diagonal slice (points) at the peak maximum and a $\sqrt{\textrm{Lorentzian}}$ fit (line) are shown in the inset to (a).}
\label{fig2}
\end{figure}

Normalized amplitude spectra are shown in Figs. \ref{fig2}(a) and \ref{fig2}(b) for the highest excitation intensity using co- (HHHH) and cross- (HVVH) linearly polarized excitation sequences, respectively, where the polarization is designated as that of pulses $A$, $B$, $C$ and the detected signal.  Because pulse $A$ is incident on the sample first and the signal is measured along $\textbf{\textit{k}}_{\textit{s}}$, coherences oscillate at negative frequencies during $\tau$ with respect to oscillations during $t$, thus the spectra are plotted as a function of negative absorption energy $\hbar\omega_{\tau}$ and positive emission energy $\hbar\omega_{t}$ along the vertical and horizontal axes, respectively.  At 200 W cm$^{-2}$, we estimate an average of 0.1 excitons are excited per QD.  Both spectra feature a peak on the diagonal, and the maximum amplitude of this peak decreases by two orders of magnitude when switching from HHHH to HVVH polarization.  In a strongly inhomogeneously broadened system, the linewidth of a slice taken along the diagonal provides the inhomogeneous linewidth, $\Gamma_{\textrm{inhom}}$, but is limited by the excitation spectrum in these experiments, whereas a slice perpendicular to the diagonal, shown in the inset of Fig. \ref{fig2}(a), is a measure of the homogeneous lineshape and the half-width half-maximum \cite{Siemens2010} (HWHM) gives the Lorentzian zero-phonon line (ZPL) width, $\gamma_{\textrm{hom}}$, at low temperature \cite{Moody2011b}.  For HVVH polarization, additional peaks appear shifted below and above the diagonal line by equal energy.

The states contributing to the features in Fig. \ref{fig2} are identified by exploiting the dipole selection rules shown in Fig. \ref{fig1}(c).  For HHHH polarization, excitons and biexcitons are optically accessible in neutral QDs, whereas in QDs containing a resident hole, Pauli blocking inhibits excitation of these excitonic states and instead positively-charged trions can form.  Because of inhomogeneity, the exciton and trion distributions overlap on the diagonal and are spectrally indistinguishable.  The excitonic contribution is eliminated by using an HVVH polarization scheme for which no quantum pathways exist for the exciton.  Local field effects, two-exciton coherences and excitation-induced effects previously measured and calculated to generate a signal at the exciton energy in quantum wells for HVVH polarization are not present, verified through auxiliary measurements similar to negative delay two-pulse FWM experiments \cite{Karaiskaj2010}.  Therefore, the remaining spectral amplitude on the diagonal in Fig. \ref{fig2}(b) isolates the trion response.  In charged QDs, optical excitation creates a single electron-hole pair, and the three-particle state forms a singlet trion. The fraction of charged QDs is estimated by noting that the third-order polarization and the ZPL width are proportional to the fourth and second power of the dipole moment, respectively.  We assume that a single dipole moment characterizes the quantum pathways leading to each multi-particle state and that the linewidths are radiatively limited.  By measuring the amplitudes and linewidths for both polarizations at low power (in the $\chi^{\textrm{(3)}}$ regime) and relating them through the dipole moments, we estimate that $52\%$ of the QDs are charged.

\begin{figure}[h]
\centering
\includegraphics[width=0.9\columnwidth]{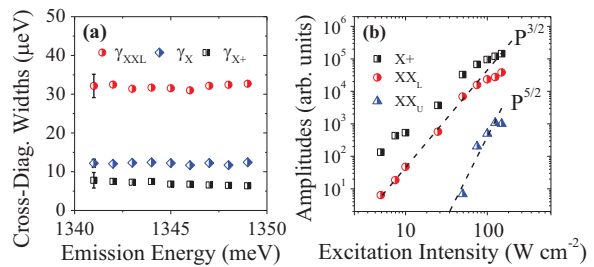}
\caption{(Color online) (a) The exciton ($\gamma_{\textrm{X}}$, diamonds), trion ($\gamma_{\textrm{X+}}$, squares) and biexciton ($\gamma_{\textrm{XX}_{\textrm{L}}}$, circles) cross-diagonal linewidths (HWHM) are shown within the ground state inhomogeneous distribution and are nearly constant for all emission energies. Representative error bars are indicated. (b) For the HVVH spectrum, a cubic power dependence of the peak amplitudes below (XX$_\textrm{L}$) and on (X+) the diagonal confirms their $\chi^{(3)}$ nature, whereas the quintic power dependence of the above-diagonal peak amplitude (XX$_\textrm{U}$) indicates that this peak arises from the $\chi^{\textrm{(5)}}$ nonlinear response amplitude radiating along $\textbf{\textit{k}}_{\textit{s}}$.  The cubic (P$^{\textrm{3/2}}$) and quintic (P$^{\textrm{5/2}}$) power dependences are indicated by the dashed lines.}
\label{fig3}
\end{figure}

Identification of the diagonal peak in Fig. \ref{fig2}(b) as a trion is further supported by a narrower ZPL width for HVVH polarization by a factor of 1.5, shown in Fig. \ref{fig3}(a), and is consistent with time-resolved photoluminescence measurements of a similar sample showing positive trions having a longer radiative lifetime than excitons \cite{Dalgarno2008}.  The cross-diagonal linewidths (HWHM) extracted from the fits, shown in Fig. \ref{fig3}(a) for the exciton ($\gamma_{\textrm{X}}$), trion ($\gamma_{\textrm{X+}}$) and biexciton ($\gamma_{\textrm{XX}_{\textrm{L}}}$), are constant within the inhomogeneous distribution FWHM for both polarization schemes and are equal to $12\pm1$, $8\pm2$ and $32\pm3$ $\mu$eV, respectively, after deconvolving the spectrometer response.  Furthermore, for all excitation intensities used, the linewidths remain constant within the experimental uncertainties (data not shown).

As mentioned previously, the cross-diagonal linewidths (HWHM) for the exciton and trion are equal to their ZPL widths at low temperature, and the values observed here are larger than the radiative lifetime-limited linewidths reported in the literature \cite{Langbein2004,Cesari2010}.  Because of different sample preparation and high-temperature annealing of the sample investigated here, comparison of the results is difficult.  However, one possible explanation for additional pure dephasing observed here could be the presence of charge-trapping sites near the QDs, which result in a fluctuating quantum-confined Stark shift of the energy levels \cite{Berthelot2006} and would dephase the excitonic states as long as the charge fluctuation times are fast compared to the formation of the photon echo.  For the biexciton, the cross-diagonal width is equal to the dephasing rate assuming that fluctuations in the ground state $\rightarrow$ exciton transition energy are perfectly correlated with those of the exciton $\rightarrow$ biexciton.  Uncorrelated broadening would lead to a greater dephasing rate and result in a distribution of binding energies at a particular emission energy, broadening the cross-diagonal linewidth.

Suppression of the exciton for HVVH polarization reveals the biexciton, which is three orders of magnitude weaker than that of the exciton in the HHHH spectrum.  Observed as beats in time-integrated FWM signals, the $\chi^{(3)}$ response of the bound biexciton has appeared in 2D rephasing spectra of GaAs quantum wells as a peak red-shifted from the diagonal along the emission energy axis $\hbar\omega_t$ by the biexciton binding energy $\Delta_{\textrm{XX}}$\cite{Bristow2009a}, similar to peak XX$_{\textrm{L}}$ observed in Fig. \ref{fig2}(b).  An additional peak, XX$_{\textrm{U}}$ in Fig. \ref{fig2}(b), appears above the diagonal red-shifted by $\Delta_{\textrm{XX}}$ along the \textit{absorption} energy axis, $\hbar\omega_{\tau}$, for excitation intensities greater than 65 W cm$^{-2}$.  A power dependence of the amplitudes for HVVH polarization, shown in Fig. \ref{fig3}(b), indicates that XX$_{\textrm{U}}$ increases with intensity as P$^{\textrm{5/2}}$ and thus is due to the $\chi^{(5)}$ biexciton nonlinear response.  This six-wave mixing signal radiates in the FWM direction only for nonlinear interactions in which either pulse $A$, $B$ or $C$ acts three times.  For excitation below 65 W cm$^{-2}$, X+ and XX$_{\textrm{L}}$ increase as P$^{\textrm{3/2}}$, confirming that the experiment is performed in the $\chi^{(3)}$ regime at low power.  The time-integrated FWM from this sample for HVVH polarization does indeed show biexciton quantum beats, but with a frequency proportional to the emission energy difference between the X+ and XX$_{\textrm{L}}$ peaks.  If strong enough, the presence of the fifth-order biexciton contribution would be distinguished in time-integrated FWM as a beat with a frequency corresponding to the energy difference between XX$_{\textrm{L}}$ and XX$_{\textrm{U}}$ \cite{Albrecht1996}.  We demonstrate here that while too weak to be observed in one-dimensional FWM experiments, certain quantum pathways of the $\chi^{\textrm{(5)}}$ biexciton nonlinear response from InAs QDs are clearly and unambiguously identified using 2DFTS.  Further separation of the $\chi^{\textrm{(5)}}$ pathways could be accomplished by using a variation of nonlinear spectroscopy that isolates the six-wave mixing signal \cite{Turner2010} in combination with higher-order Fourier-transform spectroscopy \cite{Li2012}.

In conclusion, we present 2DFT spectra of excitons, biexcitons and trions in an InAs QD ensemble for increasing excitation intensities that drive the system from the $\chi^{\textrm{(3)}}$ to the $\chi^{\textrm{(5)}}$ regime.   We find that the linewidths are independent of emission energy and that excitation-induced dephasing is absent for the excitation intensities used.  Under strong excitation, the $\chi^{\textrm{(5)}}$ biexciton nonlinear response appears in the 2D spectra as an additional peak.  While the $\chi^{\textrm{(5)}}$ biexciton contribution is too weak to be observed in the time-integrated FWM signal, it is clearly separated and resolved using 2D spectroscopy.

This work was financially supported by the Chemical Sciences, Geosciences, and Energy Biosciences Division, Office of Basic Energy Science, Office of Science, US Department of Energy, the NSF and the Deutsche Forschungsgemeinschaft.

\bibliography{RefList}{}

\begin{thebibliography}{35}%
\makeatletter
\providecommand \@ifxundefined [1]{%
 \@ifx{#1\undefined}
}%
\providecommand \@ifnum [1]{%
 \ifnum #1\expandafter \@firstoftwo
 \else \expandafter \@secondoftwo
 \fi
}%
\providecommand \@ifx [1]{%
 \ifx #1\expandafter \@firstoftwo
 \else \expandafter \@secondoftwo
 \fi
}%
\providecommand \natexlab [1]{#1}%
\providecommand \enquote  [1]{``#1''}%
\providecommand \bibnamefont  [1]{#1}%
\providecommand \bibfnamefont [1]{#1}%
\providecommand \citenamefont [1]{#1}%
\providecommand \href@noop [0]{\@secondoftwo}%
\providecommand \href [0]{\begingroup \@sanitize@url \@href}%
\providecommand \@href[1]{\@@startlink{#1}\@@href}%
\providecommand \@@href[1]{\endgroup#1\@@endlink}%
\providecommand \@sanitize@url [0]{\catcode `\\12\catcode `\$12\catcode
  `\&12\catcode `\#12\catcode `\^12\catcode `\_12\catcode `\%12\relax}%
\providecommand \@@startlink[1]{}%
\providecommand \@@endlink[0]{}%
\providecommand \url  [0]{\begingroup\@sanitize@url \@url }%
\providecommand \@url [1]{\endgroup\@href {#1}{\urlprefix }}%
\providecommand \urlprefix  [0]{URL }%
\providecommand \Eprint [0]{\href }%
\providecommand \doibase [0]{http://dx.doi.org/}%
\providecommand \selectlanguage [0]{\@gobble}%
\providecommand \bibinfo  [0]{\@secondoftwo}%
\providecommand \bibfield  [0]{\@secondoftwo}%
\providecommand \translation [1]{[#1]}%
\providecommand \BibitemOpen [0]{}%
\providecommand \bibitemStop [0]{}%
\providecommand \bibitemNoStop [0]{.\EOS\space}%
\providecommand \EOS [0]{\spacefactor3000\relax}%
\providecommand \BibitemShut  [1]{\csname bibitem#1\endcsname}%
\let\auto@bib@innerbib\@empty
\bibitem [{\citenamefont {Bayer}(2003)}]{Bayer2003}%
  \BibitemOpen
  \bibfield  {author} {\bibinfo {author} {\bibfnamefont {M.}~\bibnamefont
  {Bayer}},\ }in\ \href {http://dx.doi.org/10.1007/978-3-540-39180-7_3} {\emph
  {\bibinfo {booktitle} {Topics in Applied Physics}}},\ Vol.~\bibinfo {volume}
  {90}\ (\bibinfo  {publisher} {Springer Berlin / Heidelberg},\ \bibinfo {year}
  {2003})\ pp.\ \bibinfo {pages} {93--146}\BibitemShut {NoStop}%
\bibitem [{\citenamefont {Borri}\ \emph {et~al.}(2001)\citenamefont {Borri},
  \citenamefont {Langbein}, \citenamefont {Schneider}, \citenamefont {Woggon},
  \citenamefont {Sellin}, \citenamefont {Ouyang},\ and\ \citenamefont
  {Bimberg}}]{Borri2001}%
  \BibitemOpen
  \bibfield  {author} {\bibinfo {author} {\bibfnamefont {P.}~\bibnamefont
  {Borri}}, \bibinfo {author} {\bibfnamefont {W.}~\bibnamefont {Langbein}},
  \bibinfo {author} {\bibfnamefont {S.}~\bibnamefont {Schneider}}, \bibinfo
  {author} {\bibfnamefont {U.}~\bibnamefont {Woggon}}, \bibinfo {author}
  {\bibfnamefont {R.~L.}\ \bibnamefont {Sellin}}, \bibinfo {author}
  {\bibfnamefont {D.}~\bibnamefont {Ouyang}}, \ and\ \bibinfo {author}
  {\bibfnamefont {D.}~\bibnamefont {Bimberg}},\ }\href
  {http://link.aps.org/doi/10.1103/PhysRevLett.87.157401} {\bibfield  {journal}
  {\bibinfo  {journal} {Phys. Rev. Lett.}\ }\textbf {\bibinfo {volume} {87}},\
  \bibinfo {pages} {157401} (\bibinfo {year} {2001})}\BibitemShut {NoStop}%
\bibitem [{\citenamefont {Berry}\ \emph {et~al.}(2006)\citenamefont {Berry},
  \citenamefont {Stevens}, \citenamefont {Mirin},\ and\ \citenamefont
  {Silverman}}]{Berry2006}%
  \BibitemOpen
  \bibfield  {author} {\bibinfo {author} {\bibfnamefont {J.~J.}\ \bibnamefont
  {Berry}}, \bibinfo {author} {\bibfnamefont {M.~J.}\ \bibnamefont {Stevens}},
  \bibinfo {author} {\bibfnamefont {R.~P.}\ \bibnamefont {Mirin}}, \ and\
  \bibinfo {author} {\bibfnamefont {K.~L.}\ \bibnamefont {Silverman}},\ }\href
  {http://dx.doi.org/10.1063/1.2172291} {\bibfield  {journal} {\bibinfo
  {journal} {Appl. Phys. Lett.}\ }\textbf {\bibinfo {volume} {88}},\ \bibinfo
  {pages} {061114} (\bibinfo {year} {2006})}\BibitemShut {NoStop}%
\bibitem [{\citenamefont {Brunner}(1994)}]{Brunner1994}%
  \BibitemOpen
  \bibfield  {author} {\bibinfo {author} {\bibfnamefont {K.}~\bibnamefont
  {Brunner}},\ }\href {http://dx.doi.org/10.1063/1.111265} {\bibfield
  {journal} {\bibinfo  {journal} {Applied Physics Letters}\ }\textbf {\bibinfo
  {volume} {64}},\ \bibinfo {pages} {3320} (\bibinfo {year}
  {1994})}\BibitemShut {NoStop}%
\bibitem [{\citenamefont {Gammon}\ \emph {et~al.}(1995)\citenamefont {Gammon},
  \citenamefont {Snow},\ and\ \citenamefont {Katzer}}]{Gammon1995}%
  \BibitemOpen
  \bibfield  {author} {\bibinfo {author} {\bibfnamefont {D.}~\bibnamefont
  {Gammon}}, \bibinfo {author} {\bibfnamefont {E.~S.}\ \bibnamefont {Snow}}, \
  and\ \bibinfo {author} {\bibfnamefont {D.~S.}\ \bibnamefont {Katzer}},\
  }\href {http://dx.doi.org/10.1063/1.114557} {\bibfield  {journal} {\bibinfo
  {journal} {Appl. Phys. Lett.}\ }\textbf {\bibinfo {volume} {67}},\ \bibinfo
  {pages} {2391} (\bibinfo {year} {1995})}\BibitemShut {NoStop}%
\bibitem [{\citenamefont {Rodt}\ \emph {et~al.}(2002)\citenamefont {Rodt},
  \citenamefont {Schliwa}, \citenamefont {Heitz}, \citenamefont
  {T$\ddot{\textrm{u}}$rck}, \citenamefont {Stier}, \citenamefont {Sellin},
  \citenamefont {Strassburg}, \citenamefont {Pohl},\ and\ \citenamefont
  {Bimberg}}]{Rodt2002}%
  \BibitemOpen
  \bibfield  {author} {\bibinfo {author} {\bibfnamefont {S.}~\bibnamefont
  {Rodt}}, \bibinfo {author} {\bibfnamefont {A.}~\bibnamefont {Schliwa}},
  \bibinfo {author} {\bibfnamefont {R.}~\bibnamefont {Heitz}}, \bibinfo
  {author} {\bibfnamefont {V.}~\bibnamefont {T$\ddot{\textrm{u}}$rck}},
  \bibinfo {author} {\bibfnamefont {O.}~\bibnamefont {Stier}}, \bibinfo
  {author} {\bibfnamefont {R.~L.}\ \bibnamefont {Sellin}}, \bibinfo {author}
  {\bibfnamefont {M.}~\bibnamefont {Strassburg}}, \bibinfo {author}
  {\bibfnamefont {U.~W.}\ \bibnamefont {Pohl}}, \ and\ \bibinfo {author}
  {\bibfnamefont {D.}~\bibnamefont {Bimberg}},\ }\href
  {http://dx.doi.org/10.1002/1521-3951(200211)234:1<354::AID-PSSB354>3.0.CO;2-N}
  {\bibfield  {journal} {\bibinfo  {journal} {Phys. Status Solidi B}\ }\textbf
  {\bibinfo {volume} {234}},\ \bibinfo {pages} {354} (\bibinfo {year}
  {2002})}\BibitemShut {NoStop}%
\bibitem [{\citenamefont {Gammon}\ \emph {et~al.}(1996)\citenamefont {Gammon},
  \citenamefont {Snow}, \citenamefont {Shanabrook}, \citenamefont {Katzer},\
  and\ \citenamefont {Park}}]{Gammon1996a}%
  \BibitemOpen
  \bibfield  {author} {\bibinfo {author} {\bibfnamefont {D.}~\bibnamefont
  {Gammon}}, \bibinfo {author} {\bibfnamefont {E.~S.}\ \bibnamefont {Snow}},
  \bibinfo {author} {\bibfnamefont {B.~V.}\ \bibnamefont {Shanabrook}},
  \bibinfo {author} {\bibfnamefont {D.~S.}\ \bibnamefont {Katzer}}, \ and\
  \bibinfo {author} {\bibfnamefont {D.}~\bibnamefont {Park}},\ }\href {\doibase
  10.1126/science.273.5271.87} {\bibfield  {journal} {\bibinfo  {journal}
  {Science}\ }\textbf {\bibinfo {volume} {273}},\ \bibinfo {pages} {87}
  (\bibinfo {year} {1996})}\BibitemShut {NoStop}%
\bibitem [{\citenamefont {Besombes}\ \emph {et~al.}(2001)\citenamefont
  {Besombes}, \citenamefont {Kheng}, \citenamefont {Marsal},\ and\
  \citenamefont {Mariette}}]{Besombes2001}%
  \BibitemOpen
  \bibfield  {author} {\bibinfo {author} {\bibfnamefont {L.}~\bibnamefont
  {Besombes}}, \bibinfo {author} {\bibfnamefont {K.}~\bibnamefont {Kheng}},
  \bibinfo {author} {\bibfnamefont {L.}~\bibnamefont {Marsal}}, \ and\ \bibinfo
  {author} {\bibfnamefont {H.}~\bibnamefont {Mariette}},\ }\href
  {http://link.aps.org/doi/10.1103/PhysRevB.63.155307} {\bibfield  {journal}
  {\bibinfo  {journal} {Phys. Rev. B}\ }\textbf {\bibinfo {volume} {63}},\
  \bibinfo {pages} {155307} (\bibinfo {year} {2001})}\BibitemShut {NoStop}%
\bibitem [{\citenamefont {Ortner}\ \emph {et~al.}(2004)\citenamefont {Ortner},
  \citenamefont {Yakovlev}, \citenamefont {Bayer}, \citenamefont {Rudin},
  \citenamefont {Reinecke}, \citenamefont {Fafard}, \citenamefont
  {Wasilewski},\ and\ \citenamefont {Forchel}}]{Ortner2004}%
  \BibitemOpen
  \bibfield  {author} {\bibinfo {author} {\bibfnamefont {G.}~\bibnamefont
  {Ortner}}, \bibinfo {author} {\bibfnamefont {D.~R.}\ \bibnamefont
  {Yakovlev}}, \bibinfo {author} {\bibfnamefont {M.}~\bibnamefont {Bayer}},
  \bibinfo {author} {\bibfnamefont {S.}~\bibnamefont {Rudin}}, \bibinfo
  {author} {\bibfnamefont {T.~L.}\ \bibnamefont {Reinecke}}, \bibinfo {author}
  {\bibfnamefont {S.}~\bibnamefont {Fafard}}, \bibinfo {author} {\bibfnamefont
  {Z.}~\bibnamefont {Wasilewski}}, \ and\ \bibinfo {author} {\bibfnamefont
  {A.}~\bibnamefont {Forchel}},\ }\href
  {http://link.aps.org/doi/10.1103/PhysRevB.70.201301} {\bibfield  {journal}
  {\bibinfo  {journal} {Phys. Rev. B}\ }\textbf {\bibinfo {volume} {70}},\
  \bibinfo {pages} {201301} (\bibinfo {year} {2004})}\BibitemShut {NoStop}%
\bibitem [{\citenamefont {Peter}\ \emph {et~al.}(2004)\citenamefont {Peter},
  \citenamefont {Hours}, \citenamefont {Senellart}, \citenamefont {Vasanelli},
  \citenamefont {Cavanna}, \citenamefont {Bloch},\ and\ \citenamefont
  {G$\acute{\textrm{e}}$rard}}]{Peter2004}%
  \BibitemOpen
  \bibfield  {author} {\bibinfo {author} {\bibfnamefont {E.}~\bibnamefont
  {Peter}}, \bibinfo {author} {\bibfnamefont {J.}~\bibnamefont {Hours}},
  \bibinfo {author} {\bibfnamefont {P.}~\bibnamefont {Senellart}}, \bibinfo
  {author} {\bibfnamefont {A.}~\bibnamefont {Vasanelli}}, \bibinfo {author}
  {\bibfnamefont {A.}~\bibnamefont {Cavanna}}, \bibinfo {author} {\bibfnamefont
  {J.}~\bibnamefont {Bloch}}, \ and\ \bibinfo {author} {\bibfnamefont {J.~M.}\
  \bibnamefont {G$\acute{\textrm{e}}$rard}},\ }\href
  {http://link.aps.org/doi/10.1103/PhysRevB.69.041307} {\bibfield  {journal}
  {\bibinfo  {journal} {Phys. Rev. B}\ }\textbf {\bibinfo {volume} {69}},\
  \bibinfo {pages} {041307} (\bibinfo {year} {2004})}\BibitemShut {NoStop}%
\bibitem [{\citenamefont {Rudin}\ \emph {et~al.}(2006)\citenamefont {Rudin},
  \citenamefont {Reinecke},\ and\ \citenamefont {Bayer}}]{Rudin2006}%
  \BibitemOpen
  \bibfield  {author} {\bibinfo {author} {\bibfnamefont {S.}~\bibnamefont
  {Rudin}}, \bibinfo {author} {\bibfnamefont {T.~L.}\ \bibnamefont {Reinecke}},
  \ and\ \bibinfo {author} {\bibfnamefont {M.}~\bibnamefont {Bayer}},\ }\href
  {http://link.aps.org/doi/10.1103/PhysRevB.74.161305} {\bibfield  {journal}
  {\bibinfo  {journal} {Phys. Rev. B}\ }\textbf {\bibinfo {volume} {74}},\
  \bibinfo {pages} {161305} (\bibinfo {year} {2006})}\BibitemShut {NoStop}%
\bibitem [{\citenamefont {Bayer~\textit{et al.}}(2002)}]{Bayer2002a}%
  \BibitemOpen
  \bibfield  {author} {\bibinfo {author} {\bibfnamefont {M.}~\bibnamefont
  {Bayer~\textit{et al.}}},\ }\href
  {http://link.aps.org/doi/10.1103/PhysRevB.65.195315} {\bibfield  {journal}
  {\bibinfo  {journal} {Phys. Rev. B}\ }\textbf {\bibinfo {volume} {65}},\
  \bibinfo {pages} {195315} (\bibinfo {year} {2002})}\BibitemShut {NoStop}%
\bibitem [{\citenamefont {Schliwa}\ \emph {et~al.}(2009)\citenamefont
  {Schliwa}, \citenamefont {Winkelnkemper},\ and\ \citenamefont
  {Bimberg}}]{Schliwa2009}%
  \BibitemOpen
  \bibfield  {author} {\bibinfo {author} {\bibfnamefont {A.}~\bibnamefont
  {Schliwa}}, \bibinfo {author} {\bibfnamefont {M.}~\bibnamefont
  {Winkelnkemper}}, \ and\ \bibinfo {author} {\bibfnamefont {D.}~\bibnamefont
  {Bimberg}},\ }\href {http://link.aps.org/doi/10.1103/PhysRevB.79.075443}
  {\bibfield  {journal} {\bibinfo  {journal} {Phys. Rev. B}\ }\textbf {\bibinfo
  {volume} {79}},\ \bibinfo {pages} {075443} (\bibinfo {year}
  {2009})}\BibitemShut {NoStop}%
\bibitem [{\citenamefont {Chemla}\ and\ \citenamefont
  {Shah}(2001)}]{Chemla2001}%
  \BibitemOpen
  \bibfield  {author} {\bibinfo {author} {\bibfnamefont {D.~S.}\ \bibnamefont
  {Chemla}}\ and\ \bibinfo {author} {\bibfnamefont {J.}~\bibnamefont {Shah}},\
  }\href {http://dx.doi.org/10.1038/35079000} {\bibfield  {journal} {\bibinfo
  {journal} {Nature}\ }\textbf {\bibinfo {volume} {411}},\ \bibinfo {pages}
  {549} (\bibinfo {year} {2001})}\BibitemShut {NoStop}%
\bibitem [{\citenamefont {Fan}\ \emph {et~al.}(1998)\citenamefont {Fan},
  \citenamefont {Takagahara}, \citenamefont {Cunningham},\ and\ \citenamefont
  {Wang}}]{Fan1998}%
  \BibitemOpen
  \bibfield  {author} {\bibinfo {author} {\bibfnamefont {X.}~\bibnamefont
  {Fan}}, \bibinfo {author} {\bibfnamefont {T.}~\bibnamefont {Takagahara}},
  \bibinfo {author} {\bibfnamefont {J.}~\bibnamefont {Cunningham}}, \ and\
  \bibinfo {author} {\bibfnamefont {H.}~\bibnamefont {Wang}},\ }\href {\doibase
  10.1016/S0038-1098(98)00461-X} {\bibfield  {journal} {\bibinfo  {journal}
  {Solid State Commun.}\ }\textbf {\bibinfo {volume} {108}},\ \bibinfo {pages}
  {857} (\bibinfo {year} {1998})}\BibitemShut {NoStop}%
\bibitem [{\citenamefont {Langbein}\ \emph {et~al.}(2004)\citenamefont
  {Langbein}, \citenamefont {Borri}, \citenamefont {Woggon}, \citenamefont
  {Stavarache}, \citenamefont {Reuter},\ and\ \citenamefont
  {Wieck}}]{Langbein2004}%
  \BibitemOpen
  \bibfield  {author} {\bibinfo {author} {\bibfnamefont {W.}~\bibnamefont
  {Langbein}}, \bibinfo {author} {\bibfnamefont {P.}~\bibnamefont {Borri}},
  \bibinfo {author} {\bibfnamefont {U.}~\bibnamefont {Woggon}}, \bibinfo
  {author} {\bibfnamefont {V.}~\bibnamefont {Stavarache}}, \bibinfo {author}
  {\bibfnamefont {D.}~\bibnamefont {Reuter}}, \ and\ \bibinfo {author}
  {\bibfnamefont {A.~D.}\ \bibnamefont {Wieck}},\ }\href
  {http://link.aps.org/doi/10.1103/PhysRevB.70.033301} {\bibfield  {journal}
  {\bibinfo  {journal} {Phys. Rev. B}\ }\textbf {\bibinfo {volume} {70}},\
  \bibinfo {pages} {033301} (\bibinfo {year} {2004})}\BibitemShut {NoStop}%
\bibitem [{\citenamefont {Borri}\ and\ \citenamefont
  {Langbein}(2007)}]{Borri2007}%
  \BibitemOpen
  \bibfield  {author} {\bibinfo {author} {\bibfnamefont {P.}~\bibnamefont
  {Borri}}\ and\ \bibinfo {author} {\bibfnamefont {W.}~\bibnamefont
  {Langbein}},\ }\href {http://stacks.iop.org/0953-8984/19/i=29/a=295201}
  {\bibfield  {journal} {\bibinfo  {journal} {J. Phys.: Condens. Matter}\
  }\textbf {\bibinfo {volume} {19}},\ \bibinfo {pages} {295201} (\bibinfo
  {year} {2007})}\BibitemShut {NoStop}%
\bibitem [{\citenamefont {Cesari}\ \emph {et~al.}(2010)\citenamefont {Cesari},
  \citenamefont {Langbein},\ and\ \citenamefont {Borri}}]{Cesari2010}%
  \BibitemOpen
  \bibfield  {author} {\bibinfo {author} {\bibfnamefont {V.}~\bibnamefont
  {Cesari}}, \bibinfo {author} {\bibfnamefont {W.}~\bibnamefont {Langbein}}, \
  and\ \bibinfo {author} {\bibfnamefont {P.}~\bibnamefont {Borri}},\ }\href
  {http://link.aps.org/doi/10.1103/PhysRevB.82.195314} {\bibfield  {journal}
  {\bibinfo  {journal} {Phys. Rev. B}\ }\textbf {\bibinfo {volume} {82}},\
  \bibinfo {pages} {195314} (\bibinfo {year} {2010})}\BibitemShut {NoStop}%
\bibitem [{\citenamefont {Cundiff}\ \emph {et~al.}(2012)\citenamefont
  {Cundiff}, \citenamefont {Bristow}, \citenamefont {Siemens}, \citenamefont
  {Li}, \citenamefont {Moody}, \citenamefont {Karaiskaj}, \citenamefont {Dai},\
  and\ \citenamefont {Zhang}}]{Cundiff2012}%
  \BibitemOpen
  \bibfield  {author} {\bibinfo {author} {\bibfnamefont {S.~T.}\ \bibnamefont
  {Cundiff}}, \bibinfo {author} {\bibfnamefont {A.~D.}\ \bibnamefont
  {Bristow}}, \bibinfo {author} {\bibfnamefont {M.}~\bibnamefont {Siemens}},
  \bibinfo {author} {\bibfnamefont {H.}~\bibnamefont {Li}}, \bibinfo {author}
  {\bibfnamefont {G.}~\bibnamefont {Moody}}, \bibinfo {author} {\bibfnamefont
  {D.}~\bibnamefont {Karaiskaj}}, \bibinfo {author} {\bibfnamefont
  {X.}~\bibnamefont {Dai}}, \ and\ \bibinfo {author} {\bibfnamefont
  {T.}~\bibnamefont {Zhang}},\ }\href@noop {} {\bibfield  {journal} {\bibinfo
  {journal} {IEEE J. Sel. Topics Quantum Electr.}\ }\textbf {\bibinfo {volume}
  {18}},\ \bibinfo {pages} {318} (\bibinfo {year} {2012})}\BibitemShut
  {NoStop}%
\bibitem [{\citenamefont {Kasprzak}\ \emph {et~al.}(2011)\citenamefont
  {Kasprzak}, \citenamefont {Patton}, \citenamefont {Savona},\ and\
  \citenamefont {Langbein}}]{KasprzakJ.2011}%
  \BibitemOpen
  \bibfield  {author} {\bibinfo {author} {\bibfnamefont {J.}~\bibnamefont
  {Kasprzak}}, \bibinfo {author} {\bibfnamefont {B.}~\bibnamefont {Patton}},
  \bibinfo {author} {\bibfnamefont {V.}~\bibnamefont {Savona}}, \ and\ \bibinfo
  {author} {\bibfnamefont {W.}~\bibnamefont {Langbein}},\ }\href
  {http://dx.doi.org/10.1038/nphoton.2010.284} {\bibfield  {journal} {\bibinfo
  {journal} {Nat. Photonics}\ }\textbf {\bibinfo {volume} {5}},\ \bibinfo
  {pages} {123} (\bibinfo {year} {2011})}\BibitemShut {NoStop}%
\bibitem [{\citenamefont {Kasprzak}\ and\ \citenamefont
  {Langbein}(2012)}]{Kasprzak2012}%
  \BibitemOpen
  \bibfield  {author} {\bibinfo {author} {\bibfnamefont {J.}~\bibnamefont
  {Kasprzak}}\ and\ \bibinfo {author} {\bibfnamefont {W.}~\bibnamefont
  {Langbein}},\ }\href {http://josab.osa.org/abstract.cfm?URI=josab-29-7-1766}
  {\bibfield  {journal} {\bibinfo  {journal} {J. Opt. Soc. Am. B}\ }\textbf
  {\bibinfo {volume} {29}},\ \bibinfo {pages} {1766} (\bibinfo {year}
  {2012})}\BibitemShut {NoStop}%
\bibitem [{\citenamefont {Moody}\ \emph
  {et~al.}(2011{\natexlab{a}})\citenamefont {Moody}, \citenamefont {Siemens},
  \citenamefont {Bristow}, \citenamefont {Dai}, \citenamefont {Karaiskaj},
  \citenamefont {Bracker}, \citenamefont {Gammon},\ and\ \citenamefont
  {Cundiff}}]{Moody2011b}%
  \BibitemOpen
  \bibfield  {author} {\bibinfo {author} {\bibfnamefont {G.}~\bibnamefont
  {Moody}}, \bibinfo {author} {\bibfnamefont {M.~E.}\ \bibnamefont {Siemens}},
  \bibinfo {author} {\bibfnamefont {A.~D.}\ \bibnamefont {Bristow}}, \bibinfo
  {author} {\bibfnamefont {X.}~\bibnamefont {Dai}}, \bibinfo {author}
  {\bibfnamefont {D.}~\bibnamefont {Karaiskaj}}, \bibinfo {author}
  {\bibfnamefont {A.~S.}\ \bibnamefont {Bracker}}, \bibinfo {author}
  {\bibfnamefont {D.}~\bibnamefont {Gammon}}, \ and\ \bibinfo {author}
  {\bibfnamefont {S.~T.}\ \bibnamefont {Cundiff}},\ }\href
  {http://link.aps.org/doi/10.1103/PhysRevB.83.115324} {\bibfield  {journal}
  {\bibinfo  {journal} {Phys. Rev. B}\ }\textbf {\bibinfo {volume} {83}},\
  \bibinfo {pages} {115324} (\bibinfo {year} {2011}{\natexlab{a}})}\BibitemShut
  {NoStop}%
\bibitem [{\citenamefont {Moody}\ \emph
  {et~al.}(2011{\natexlab{b}})\citenamefont {Moody}, \citenamefont {Siemens},
  \citenamefont {Bristow}, \citenamefont {Dai}, \citenamefont {Bracker},
  \citenamefont {Gammon},\ and\ \citenamefont {Cundiff}}]{Moody2011}%
  \BibitemOpen
  \bibfield  {author} {\bibinfo {author} {\bibfnamefont {G.}~\bibnamefont
  {Moody}}, \bibinfo {author} {\bibfnamefont {M.~E.}\ \bibnamefont {Siemens}},
  \bibinfo {author} {\bibfnamefont {A.~D.}\ \bibnamefont {Bristow}}, \bibinfo
  {author} {\bibfnamefont {X.}~\bibnamefont {Dai}}, \bibinfo {author}
  {\bibfnamefont {A.~S.}\ \bibnamefont {Bracker}}, \bibinfo {author}
  {\bibfnamefont {D.}~\bibnamefont {Gammon}}, \ and\ \bibinfo {author}
  {\bibfnamefont {S.~T.}\ \bibnamefont {Cundiff}},\ }\href
  {http://link.aps.org/doi/10.1103/PhysRevB.83.245316} {\bibfield  {journal}
  {\bibinfo  {journal} {Phys. Rev. B}\ }\textbf {\bibinfo {volume} {83}},\
  \bibinfo {pages} {245316} (\bibinfo {year} {2011}{\natexlab{b}})}\BibitemShut
  {NoStop}%
\bibitem [{\citenamefont {Leon}\ \emph {et~al.}(1996)\citenamefont {Leon},
  \citenamefont {Kim}, \citenamefont {Jagadish}, \citenamefont {Gal},
  \citenamefont {Zou},\ and\ \citenamefont {Cockayne}}]{Leon1996}%
  \BibitemOpen
  \bibfield  {author} {\bibinfo {author} {\bibfnamefont {R.}~\bibnamefont
  {Leon}}, \bibinfo {author} {\bibfnamefont {Y.}~\bibnamefont {Kim}}, \bibinfo
  {author} {\bibfnamefont {C.}~\bibnamefont {Jagadish}}, \bibinfo {author}
  {\bibfnamefont {M.}~\bibnamefont {Gal}}, \bibinfo {author} {\bibfnamefont
  {J.}~\bibnamefont {Zou}}, \ and\ \bibinfo {author} {\bibfnamefont {D.~J.~H.}\
  \bibnamefont {Cockayne}},\ }\href {http://dx.doi.org/10.1063/1.117467}
  {\bibfield  {journal} {\bibinfo  {journal} {Appl. Phys. Lett.}\ }\textbf
  {\bibinfo {volume} {69}},\ \bibinfo {pages} {1888} (\bibinfo {year}
  {1996})}\BibitemShut {NoStop}%
\bibitem [{\citenamefont {Greilich~\textit{et al.}}(2006)}]{Greilich2006}%
  \BibitemOpen
  \bibfield  {author} {\bibinfo {author} {\bibfnamefont {A.}~\bibnamefont
  {Greilich~\textit{et al.}}},\ }\href@noop {} {\bibfield  {journal} {\bibinfo
  {journal} {Phys. Rev. B}\ }\textbf {\bibinfo {volume} {73}},\ \bibinfo
  {pages} {045323} (\bibinfo {year} {2006})}\BibitemShut {NoStop}%
\bibitem [{\citenamefont {Moody}\ \emph {et~al.}(2012)\citenamefont {Moody},
  \citenamefont {Singh}, \citenamefont {Li}, \citenamefont {Akimov},
  \citenamefont {Bayer}, \citenamefont {Reuter}, \citenamefont {Wieck},\ and\
  \citenamefont {Cundiff}}]{Moody2012b}%
  \BibitemOpen
  \bibfield  {author} {\bibinfo {author} {\bibfnamefont {G.}~\bibnamefont
  {Moody}}, \bibinfo {author} {\bibfnamefont {R.}~\bibnamefont {Singh}},
  \bibinfo {author} {\bibfnamefont {H.}~\bibnamefont {Li}}, \bibinfo {author}
  {\bibfnamefont {I.~A.}\ \bibnamefont {Akimov}}, \bibinfo {author}
  {\bibfnamefont {M.}~\bibnamefont {Bayer}}, \bibinfo {author} {\bibfnamefont
  {D.}~\bibnamefont {Reuter}}, \bibinfo {author} {\bibfnamefont
  {A.}~\bibnamefont {Wieck}}, \ and\ \bibinfo {author} {\bibfnamefont {S.~T.}\
  \bibnamefont {Cundiff}},\ }\href@noop {} {\bibfield  {journal} {\bibinfo
  {journal} {\textit{submitted for publication}}\ } (\bibinfo {year}
  {2012})}\BibitemShut {NoStop}%
\bibitem [{\citenamefont {Bristow}\ \emph
  {et~al.}(2009{\natexlab{a}})\citenamefont {Bristow}, \citenamefont
  {Karaiskaj}, \citenamefont {Dai}, \citenamefont {Zhang}, \citenamefont
  {Carlsson}, \citenamefont {Hagen}, \citenamefont {Jimenez},\ and\
  \citenamefont {Cundiff}}]{Bristow2009}%
  \BibitemOpen
  \bibfield  {author} {\bibinfo {author} {\bibfnamefont {A.~D.}\ \bibnamefont
  {Bristow}}, \bibinfo {author} {\bibfnamefont {D.}~\bibnamefont {Karaiskaj}},
  \bibinfo {author} {\bibfnamefont {X.}~\bibnamefont {Dai}}, \bibinfo {author}
  {\bibfnamefont {T.}~\bibnamefont {Zhang}}, \bibinfo {author} {\bibfnamefont
  {C.}~\bibnamefont {Carlsson}}, \bibinfo {author} {\bibfnamefont {K.~R.}\
  \bibnamefont {Hagen}}, \bibinfo {author} {\bibfnamefont {R.}~\bibnamefont
  {Jimenez}}, \ and\ \bibinfo {author} {\bibfnamefont {S.~T.}\ \bibnamefont
  {Cundiff}},\ }\href {http://dx.doi.org/10.1063/1.3184103} {\bibfield
  {journal} {\bibinfo  {journal} {Rev. Sci. Instrum.}\ }\textbf {\bibinfo
  {volume} {80}},\ \bibinfo {pages} {073108} (\bibinfo {year}
  {2009}{\natexlab{a}})}\BibitemShut {NoStop}%
\bibitem [{\citenamefont {Siemens}\ \emph {et~al.}(2010)\citenamefont
  {Siemens}, \citenamefont {Moody}, \citenamefont {Li}, \citenamefont
  {Bristow},\ and\ \citenamefont {Cundiff}}]{Siemens2010}%
  \BibitemOpen
  \bibfield  {author} {\bibinfo {author} {\bibfnamefont {M.~E.}\ \bibnamefont
  {Siemens}}, \bibinfo {author} {\bibfnamefont {G.}~\bibnamefont {Moody}},
  \bibinfo {author} {\bibfnamefont {H.}~\bibnamefont {Li}}, \bibinfo {author}
  {\bibfnamefont {A.~D.}\ \bibnamefont {Bristow}}, \ and\ \bibinfo {author}
  {\bibfnamefont {S.~T.}\ \bibnamefont {Cundiff}},\ }\href
  {http://www.opticsexpress.org/abstract.cfm?URI=oe-18-17-17699} {\bibfield
  {journal} {\bibinfo  {journal} {Opt. Express}\ }\textbf {\bibinfo {volume}
  {18}},\ \bibinfo {pages} {17699} (\bibinfo {year} {2010})}\BibitemShut
  {NoStop}%
\bibitem [{\citenamefont {Karaiskaj}\ \emph {et~al.}(2010)\citenamefont
  {Karaiskaj}, \citenamefont {Bristow}, \citenamefont {Yang}, \citenamefont
  {Dai}, \citenamefont {Mirin}, \citenamefont {Mukamel},\ and\ \citenamefont
  {Cundiff}}]{Karaiskaj2010}%
  \BibitemOpen
  \bibfield  {author} {\bibinfo {author} {\bibfnamefont {D.}~\bibnamefont
  {Karaiskaj}}, \bibinfo {author} {\bibfnamefont {A.~D.}\ \bibnamefont
  {Bristow}}, \bibinfo {author} {\bibfnamefont {L.}~\bibnamefont {Yang}},
  \bibinfo {author} {\bibfnamefont {X.}~\bibnamefont {Dai}}, \bibinfo {author}
  {\bibfnamefont {R.~P.}\ \bibnamefont {Mirin}}, \bibinfo {author}
  {\bibfnamefont {S.}~\bibnamefont {Mukamel}}, \ and\ \bibinfo {author}
  {\bibfnamefont {S.~T.}\ \bibnamefont {Cundiff}},\ }\href
  {http://link.aps.org/doi/10.1103/PhysRevLett.104.117401} {\bibfield
  {journal} {\bibinfo  {journal} {Phys. Rev. Lett.}\ }\textbf {\bibinfo
  {volume} {104}},\ \bibinfo {pages} {117401} (\bibinfo {year}
  {2010})}\BibitemShut {NoStop}%
\bibitem [{\citenamefont {Dalgarno}\ \emph {et~al.}(2008)\citenamefont
  {Dalgarno}, \citenamefont {Smith}, \citenamefont {McFarlane}, \citenamefont
  {Gerardot}, \citenamefont {Karrai}, \citenamefont {Badolato}, \citenamefont
  {Petroff},\ and\ \citenamefont {Warburton}}]{Dalgarno2008}%
  \BibitemOpen
  \bibfield  {author} {\bibinfo {author} {\bibfnamefont {P.~A.}\ \bibnamefont
  {Dalgarno}}, \bibinfo {author} {\bibfnamefont {J.~M.}\ \bibnamefont {Smith}},
  \bibinfo {author} {\bibfnamefont {J.}~\bibnamefont {McFarlane}}, \bibinfo
  {author} {\bibfnamefont {B.~D.}\ \bibnamefont {Gerardot}}, \bibinfo {author}
  {\bibfnamefont {K.}~\bibnamefont {Karrai}}, \bibinfo {author} {\bibfnamefont
  {A.}~\bibnamefont {Badolato}}, \bibinfo {author} {\bibfnamefont {P.~M.}\
  \bibnamefont {Petroff}}, \ and\ \bibinfo {author} {\bibfnamefont {R.~J.}\
  \bibnamefont {Warburton}},\ }\href
  {http://link.aps.org/doi/10.1103/PhysRevB.77.245311} {\bibfield  {journal}
  {\bibinfo  {journal} {Phys. Rev. B}\ }\textbf {\bibinfo {volume} {77}},\
  \bibinfo {pages} {245311} (\bibinfo {year} {2008})}\BibitemShut {NoStop}%
\bibitem [{\citenamefont {Berthelot}\ \emph {et~al.}(2006)\citenamefont
  {Berthelot}, \citenamefont {Favero}, \citenamefont {Cassabois}, \citenamefont
  {Voisin}, \citenamefont {Delalande}, \citenamefont {Roussignol},
  \citenamefont {Ferreira},\ and\ \citenamefont {Gerard}}]{Berthelot2006}%
  \BibitemOpen
  \bibfield  {author} {\bibinfo {author} {\bibfnamefont {A.}~\bibnamefont
  {Berthelot}}, \bibinfo {author} {\bibfnamefont {I.}~\bibnamefont {Favero}},
  \bibinfo {author} {\bibfnamefont {G.}~\bibnamefont {Cassabois}}, \bibinfo
  {author} {\bibfnamefont {C.}~\bibnamefont {Voisin}}, \bibinfo {author}
  {\bibfnamefont {C.}~\bibnamefont {Delalande}}, \bibinfo {author}
  {\bibfnamefont {P.}~\bibnamefont {Roussignol}}, \bibinfo {author}
  {\bibfnamefont {R.}~\bibnamefont {Ferreira}}, \ and\ \bibinfo {author}
  {\bibfnamefont {J.~M.}\ \bibnamefont {Gerard}},\ }\href
  {http://dx.doi.org/10.1038/nphys433} {\bibfield  {journal} {\bibinfo
  {journal} {Nat. Phys.}\ }\textbf {\bibinfo {volume} {2}},\ \bibinfo {pages}
  {759} (\bibinfo {year} {2006})}\BibitemShut {NoStop}%
\bibitem [{\citenamefont {Bristow}\ \emph
  {et~al.}(2009{\natexlab{b}})\citenamefont {Bristow}, \citenamefont
  {Karaiskaj}, \citenamefont {Dai}, \citenamefont {Mirin},\ and\ \citenamefont
  {Cundiff}}]{Bristow2009a}%
  \BibitemOpen
  \bibfield  {author} {\bibinfo {author} {\bibfnamefont {A.~D.}\ \bibnamefont
  {Bristow}}, \bibinfo {author} {\bibfnamefont {D.}~\bibnamefont {Karaiskaj}},
  \bibinfo {author} {\bibfnamefont {X.}~\bibnamefont {Dai}}, \bibinfo {author}
  {\bibfnamefont {R.~P.}\ \bibnamefont {Mirin}}, \ and\ \bibinfo {author}
  {\bibfnamefont {S.~T.}\ \bibnamefont {Cundiff}},\ }\href
  {http://link.aps.org/doi/10.1103/PhysRevB.79.161305} {\bibfield  {journal}
  {\bibinfo  {journal} {Phys. Rev. B}\ }\textbf {\bibinfo {volume} {79}},\
  \bibinfo {pages} {161305} (\bibinfo {year} {2009}{\natexlab{b}})}\BibitemShut
  {NoStop}%
\bibitem [{\citenamefont {Albrecht~\textit{et al.}}(1996)}]{Albrecht1996}%
  \BibitemOpen
  \bibfield  {author} {\bibinfo {author} {\bibfnamefont {T.~F.}\ \bibnamefont
  {Albrecht~\textit{et al.}}},\ }\href
  {http://link.aps.org/doi/10.1103/PhysRevB.54.4436} {\bibfield  {journal}
  {\bibinfo  {journal} {Phys. Rev. B}\ }\textbf {\bibinfo {volume} {54}},\
  \bibinfo {pages} {4436} (\bibinfo {year} {1996})}\BibitemShut {NoStop}%
\bibitem [{\citenamefont {Turner}\ and\ \citenamefont
  {Nelson}(2010)}]{Turner2010}%
  \BibitemOpen
  \bibfield  {author} {\bibinfo {author} {\bibfnamefont {D.~B.}\ \bibnamefont
  {Turner}}\ and\ \bibinfo {author} {\bibfnamefont {K.~A.}\ \bibnamefont
  {Nelson}},\ }\href {http://dx.doi.org/10.1038/nature09286} {\bibfield
  {journal} {\bibinfo  {journal} {Nature}\ }\textbf {\bibinfo {volume} {466}},\
  \bibinfo {pages} {1089} (\bibinfo {year} {2010})}\BibitemShut {NoStop}%
\bibitem [{\citenamefont {Li}\ \emph {et~al.}(2012)\citenamefont {Li},
  \citenamefont {Bristow}, \citenamefont {Siemens}, \citenamefont {Moody},\
  and\ \citenamefont {Cundiff}}]{Li2012}%
  \BibitemOpen
  \bibfield  {author} {\bibinfo {author} {\bibfnamefont {H.}~\bibnamefont
  {Li}}, \bibinfo {author} {\bibfnamefont {A.~D.}\ \bibnamefont {Bristow}},
  \bibinfo {author} {\bibfnamefont {M.~E.}\ \bibnamefont {Siemens}}, \bibinfo
  {author} {\bibfnamefont {G.}~\bibnamefont {Moody}}, \ and\ \bibinfo {author}
  {\bibfnamefont {S.~T.}\ \bibnamefont {Cundiff}},\ }\href@noop {} {\bibfield
  {journal} {\bibinfo  {journal} {\textit{submitted for publication}}\ }
  (\bibinfo {year} {2012})}\BibitemShut {NoStop}%
\end{thebibliography}%

\end{document}